\documentclass[12pt,twocolumn]{article}

\usepackage[english]{babel}
\usepackage[utf8]{inputenc}
\usepackage[T1,T2A]{fontenc}
\usepackage{cite}
\usepackage{physics}
\usepackage{mathtools}
\usepackage{xcolor}
\usepackage{graphicx}
\usepackage{dcolumn}
\usepackage{bm}

\usepackage[utf8]{inputenc}
\usepackage{amsmath}
\usepackage{amssymb}
\usepackage{xcolor}
\usepackage{rotating}
\usepackage{lscape}
\usepackage{longtable}
\usepackage{authblk} 

\usepackage{hyperref}

\usepackage[bottom=0.5in]{geometry}
\usepackage{caption}

\title{Stability of the Milky Way Satellite Galaxy Plane under the Influence of Neighbors}

\author[1,*]{S.V. Pilipenko}
\author[1]{N.R. Arakelyan}
\affil[1]{P.N. Lebedev Physical Institute RAS, Profsoyuznaya st., 84/32, 117997 Moscow, Russia}
\affil[*]{e-mail: spilipenko@asc.rssi.ru}
\date{\today}

\begin{document}

\twocolumn[
  \begin{@twocolumnfalse}
\maketitle

\begin{abstract}
Trajectories of test particles in a time-varying nonspherical gravitational potential model of our Galaxy are considered. The role of the quadrupole component of the potential, which at distances greater than 50 kpc is associated with the distribution of matter in the Galaxy's neighborhood (mainly with the influence of the galaxy M31), is studied. It is shown that perturbations of the potential created by the environment can significantly change the trajectories of particles at distances greater than 100 kpc from the Galactic center, but the magnitude of this effect depends on the still poorly known trajectory of the galaxy M31. For some variants of this trajectory, structures resembling a "thin plane" of satellite galaxies cannot exist for more than 2-3 billion years.
\end{abstract}

\vspace{.5cm}

\end{@twocolumnfalse}
]

\section{INTRODUCTION}
\label{sec:intro}
Detailed observations of satellite galaxies and globular clusters (GCs) of the Milky Way (MW) and some other nearby galaxies have revealed the presence of "thin planes" of satellites and GCs \cite{1976MNRAS.174..695L,1995MNRAS.275..429L,2005A&A...431..517K,2008ApJ...680..287M,2007MNRAS.374.1125M,2009MNRAS.394.2223M,2015MNRAS.453.1047P,2025arXiv250601459K}. In the literature, there is widespread discussion whether this constitutes a problem for our understanding of galaxy physics and cosmology, since according to some authors, such planes are too rare in cosmological numerical models \cite{2021ApJ...923..140G,2022ARep...66..359A,2024A&A...681A..73T,2025A&A...694A..66A,2015MNRAS.452.1052L,2005A&A...431..517K,2008ApJ...680..287M}. 

The satellite galaxies forming the "thin plane" mostly have distances from 50 to 240 kpc from the Galactic center. Also, six of the most distant GCs with distances greater than 50 kpc fall into this plane, which has a thickness of about 100 kpc \cite{Arakelyan18}. GCs at such large distances from the Galactic center are presumably genetically related to satellite galaxies, i.e., they could have been lost by them due to the tidal influence of the Galaxy (see, e.g., \cite{Arakelyan22} and references therein). To test this hypothesis, it is necessary to reconstruct the orbits of GCs backward in time by several billion years. Also, to study the origin of the plane of satellite galaxies, it is necessary to calculate their orbits in the past. New GAIA data provide us with information on the three-dimensional velocities of GCs and satellite galaxies, which is already being applied to reconstruct their orbits and draw conclusions about the origin of the "thin plane," see, for example, \cite{2022A&A...657A..54B,2022ApJ...940..136P,2021ApJ...916....8L,2024A&A...681A..73T}.

Recently, a hypothesis has been proposed in the literature that the "thin plane" of satellites is a short-lived phenomenon with a lifetime of 1-3 billion years. This is supported by the correlation found in work \cite{2025A&A...694A..66A} between satellites forming the plane and stars from a tidal tail formed during a collision of M31 with another galaxy 2-3 billion years ago. Also, the argument about the destruction of the plane due to dark matter subhalos \cite{2018MNRAS.473.2212F} supports the short lifetime of the plane. {In highly detailed hydrodynamical cosmological simulations from the FIRE project, mostly short-lived planes with lifetimes less than 1 billion years are observed, but in galaxies having a massive satellite (like the Large Magellanic Cloud), they can exist for 1-3 billion years \cite{Samuel21}.}

At distances greater than 50 kpc from the center, the distribution of matter surrounding our Galaxy can influence the motion of satellite galaxies and GCs. It is well known that our Galaxy is part of the Local Group, where matter is distributed inhomogeneously: at a distance of about 700 kpc from the MW there is a galaxy M31, which is supposedly more massive than ours; the Local Filament, the Local Void, and other structures are also identified in observations. Therefore, in this work, we analyze the possible influence of the nonspherical matter distribution on the accuracy of orbit reconstruction at the Galactic periphery, in particular on the stability of the "thin plane" of satellites. The orbital periods of satellites can reach several billion years, during which the matter distribution can change significantly, which also needs to be taken into account in such an analysis.

The influence of nonspherical and time-dependent gravitational potential on the motion of objects in the Galaxy has been considered in many works. However, the main focus was on relatively small distances from the center, where the bulge and disk of the Galaxy, as well as the satellite galaxy Large Magellanic Cloud, have a significant influence \cite{2021ApJ...923..140G,2022A&A...657A..54B,CorreaMagnus22,2022ApJ...933..113R,2023Galax..11...59V,Makarov23,2025ApJ...978...79B,2025Univ...11..191M}. In our work, more distant regions of the Galaxy, 50--250 kpc from the center, are considered in application to satellite galaxies and GCs.

Due to the lack of data on the distribution of the dark matter component in our immediate neighborhood, it is proposed to use cosmological simulations for such an analysis, in which constraints from observational data were imposed on the initial conditions \cite{CLUES,Libeskind20,SLOW}. Namely, measurements of distances to thousands of galaxies allow one to reconstruct their peculiar velocities, and from their spatial distribution one can restore the distribution of gravitational potential and matter density on scales of several Mpc. Then these distributions can be evolved backward in time, and by filling gaps in the data and adding random perturbations on small scales, initial conditions for simulations can be obtained. This method successfully reproduces large structures such as the local void and nearby galaxy clusters. In the HESTIA project \cite{Libeskind20}, model analogs of our environment including an analog of the Local Group were obtained in this way. About a thousand realizations of initial conditions were generated, among which those containing a galaxy group satisfying a set of criteria such as mass ratio and distance between two large galaxies were selected \cite{CLUES,Libeskind20}.

In this work, to model the Galactic potential, results of \cite{Arakelyan25} are used, where the gravitational potential of Galaxy analogs in cosmological models from the HESTIA project was analyzed. The potential was expanded into spherical harmonics (relative to the center of the Galaxy analog), and their behavior was studied, in particular the influence of M31. The coefficients of the potential expansion and trajectories of the MW and M31 from \cite{Arakelyan25} are publicly available at \url{https://github.com/spilipenko/HESITA_MW_potential}. It was shown in \cite{Arakelyan25} that the gravitational potential at distances greater than 100 kpc from the Galactic center depends significantly on the environment and is nonstationary on timescales longer than 1 billion years. The presence of an M31 analog in HESTIA simulations leads to an increase of multipole amplitudes of the potential with distance in the interval 100 kpc $< r < d_{M31}$, where $d_{M31}$ is the distance to M31. The dipole component has the largest amplitude, but its influence on test particle orbits is substantially reduced when accounting for the non-inertial reference frame connected with the Galactic center. Therefore, the largest contribution to the nonspherical potential affecting satellite galaxy motion is the next, quadrupole component. The octupole and higher harmonics contribute significantly less.

We use two different approaches to assess the influence of nonspherical and time-dependent potential on object trajectories in the outer Galaxy. In the first approach, described in Section 2, the potential perturbation is represented as a quadrupole harmonic. In Section 3, the perturbation is modeled by a moving body -- an analog of M31. Discussion and conclusions are presented in Section 4.

\section{Influence of Quadrupole Nonsphericity on Trajectories of Outer Galactic Objects}
In this section, the influence of quadrupole potential nonsphericity on trajectories of test particles moving at the Galaxy's periphery is considered. We assume the monopole component of the potential is given by the Navarro-Frenk-White (NFW) density profile \cite{NFW}, with a virial mass of $1\times10^{12}$ M$_\odot$. The quadrupole on the sphere's surface is described by five amplitudes $a_{2m}$, where $m=-2,-1,0,1,2$, but by rotating the coordinate system, these can be reduced to two: $a_{20}$ (with $l=2$ and $m=0$) and $a_{22}$ (with $l=2$ and $m=2$). The first defines the magnitude of the azimuthally symmetric component of the potential, the second defines the component symmetric with respect to three mutually perpendicular planes. This approach is convenient for describing orbit perturbations and was applied, for example, in \cite{Ivanov25} for the orbit of a binary black hole in a nonspherical star cluster.

The multipole expansion describes the angular dependence of the potential only; it must be supplemented with a radial dependence. Results of \cite{Arakelyan25} showed that quadrupole harmonic amplitudes grow with distance approximately as $r^{3/2}$ while $r < d_{M31}$ (the distance to M31). 

If the quadrupole perturbation is assumed to be created by a point mass, then by symmetry, the perturbation is described only by the harmonic with $m=0$ if the Z-axis is directed toward the perturbing body. For this reason, we considered a potential model with a fixed monopole component (NFW profile) and a single quadrupole component with $m=0$ and amplitude
\begin{equation}
    a_{20} (r/\mathrm{kpc})^{3/2}
\end{equation}
up to a distance of 700 kpc.

\begin{figure*}
    \centering
    \includegraphics[width=\linewidth]{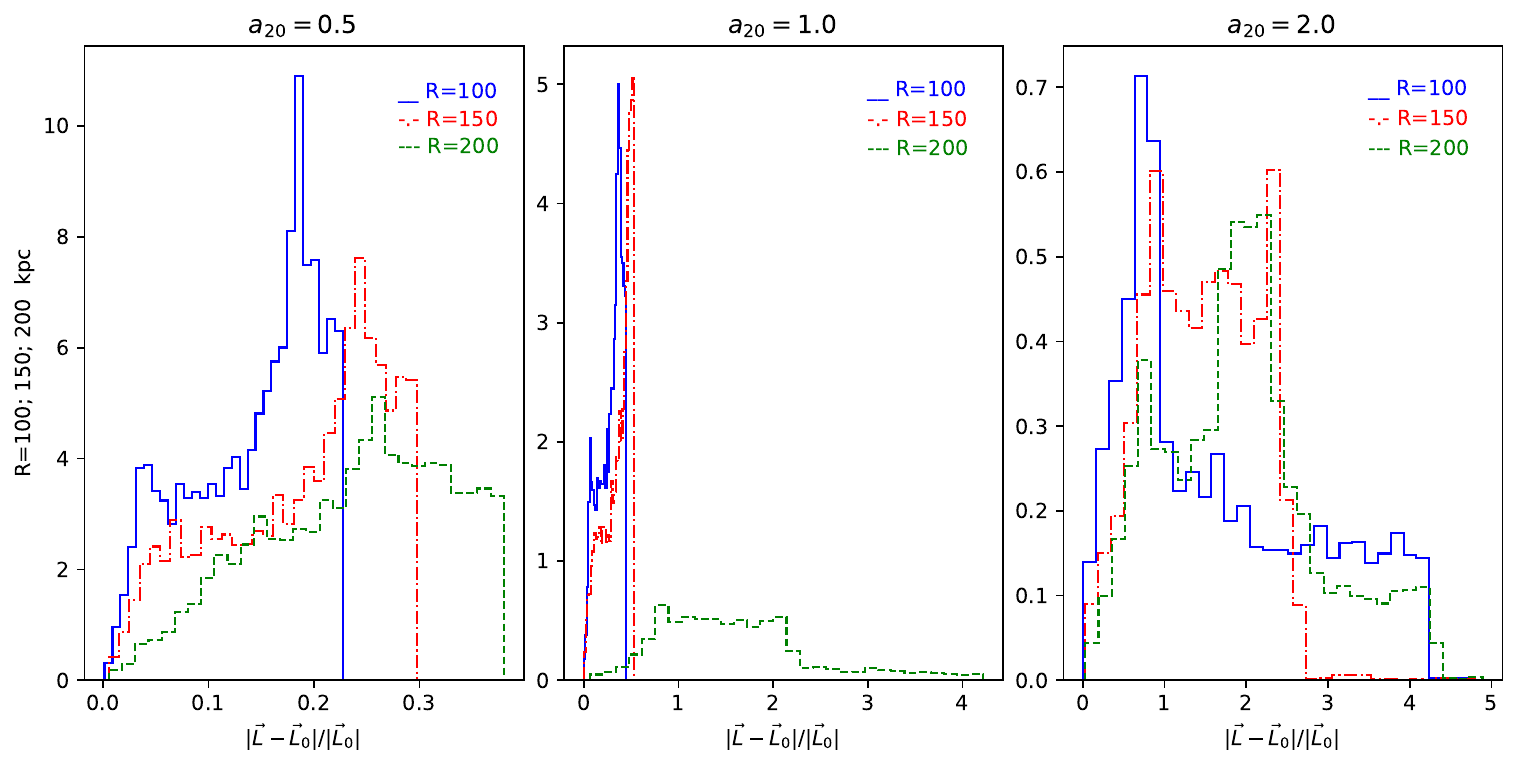}
    \caption{Histograms of the distribution of changes in angular momentum of particles initially uniformly distributed on spheres of radius 100, 150, and 200 kpc (solid, dash-dot, and dashed lines) for quadrupole component amplitudes $a_{20} = 0.5, 1.0,$ and $2.0$ (left, center, right).}
    \label{fig:lchange}
\end{figure*}

To investigate the influence of the above potential on test particle trajectories, we use the AGAMA software package \cite{2018arXiv180208255V}, which allows specifying arbitrary gravitational potentials via spherical harmonic expansions and calculating trajectories by solving equations of motion. Initial positions of test particles were randomly set on the surface of a sphere of fixed radius. To set the velocities, the circular velocities for the monopole component of the potential at this radius were calculated. Velocity directions were random but always tangential to the sphere. Thus, without quadrupole components, particles move along randomly oriented circular orbits.

To characterize changes caused by nonsphericity, the angular momentum $\vec{L}$ of particles was measured after 5 billion years, and for each particle, the length of the difference vector between the initial and final angular momentum was computed and divided by the initial angular momentum amplitude. Figure~\ref{fig:lchange} shows histograms of the resulting angular momentum differences for initial radii of 100, 150, and 200 kpc and for different quadrupole component amplitudes $a_{20}$. From this figure, one concludes that increasing the quadrupole amplitude from $a_{20}=0.5$ to $a_{20}=2.0$ substantially changes the effect of the quadrupole component on orbits. At $a_{20}=0.5$, changes in angular momentum do not exceed 30\% of the initial value at all three distances, and a special check showed that mainly the direction of $\vec{L}$ changes, while its magnitude changes by less than 10\%. At $a_{20}=1.0$, at 100 and 150 kpc, the picture is almost unchanged, but at 200 kpc, the distribution of $\vec{L}$ changes becomes much wider and can exceed 100\%. At $a_{20}=2.0$, at all three distances, the distributions become wide. Thus, exceeding a threshold quadrupole amplitude, dependent on radius, leads to significant orbit changes. It was also verified that setting the quadrupole component $a_{22}$ instead of $a_{20}$ qualitatively does not change the results.

According to data obtained in \cite{Arakelyan25}, the quadrupole amplitude in different Galaxy analog realizations is 1-1.5, close to the threshold value found here. This suggests it makes sense to solve a more complex problem by creating a more realistic model potential, which is done in the next section.

\section{Influence of Time-Dependent Nonspherical Potential on Orbital Planes}
In this section, the motion of test particles in gravitational potentials created by a two-body system — the Milky Way and M31 galaxies — is modeled. Each galaxy is modeled by an NFW density distribution with the corresponding potential. {The trajectory of M31 relative to the Galaxy analog and the masses of the two galaxies are taken from cosmological numerical models.}

\begin{figure*}
    \centering
    \includegraphics[width=0.95\linewidth]{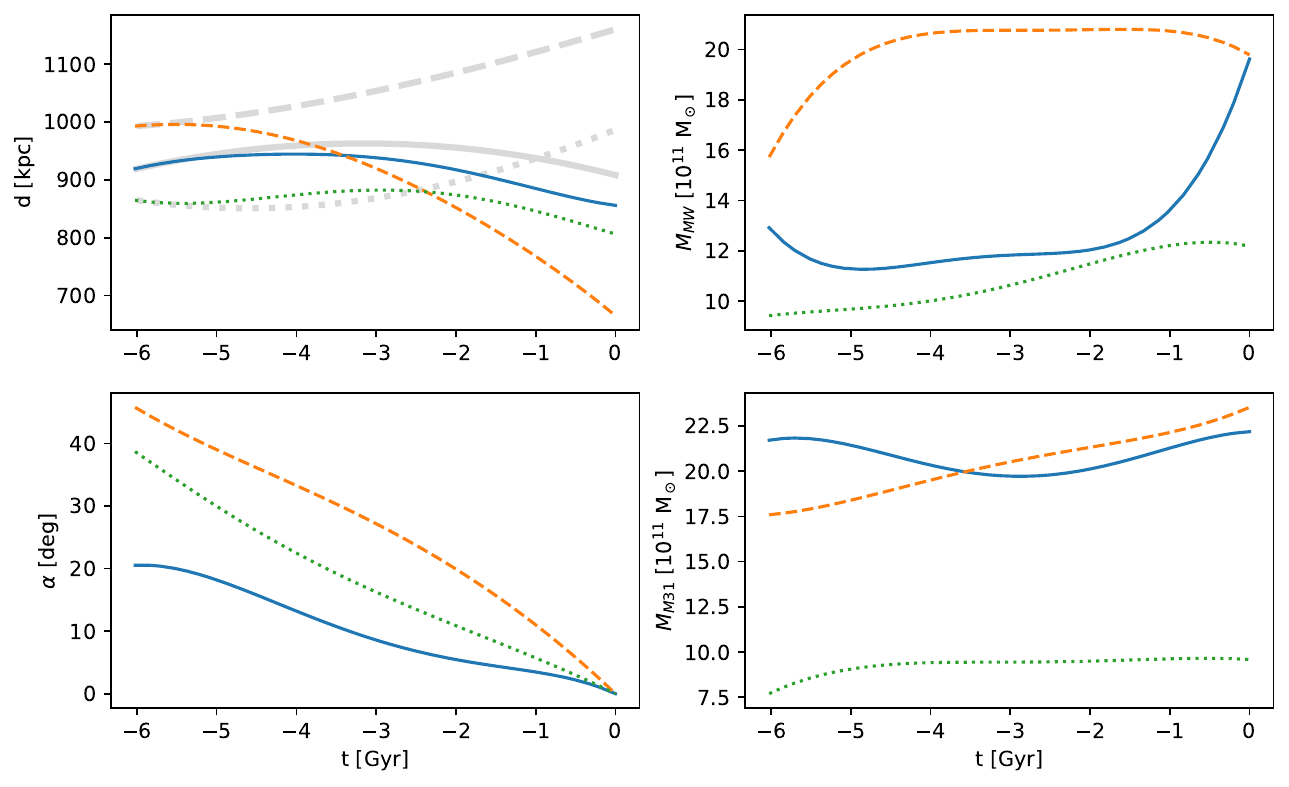}
    \caption{M31 trajectories in three model calculations. Solid line — 09\_18, dashed — 17\_11, dotted — 37\_17. Top left: distance between M31 and the Galaxy (faint thick lines show Keplerian trajectories). Bottom left: angle between the direction to M31 from the Galactic center at present (t=0) and other times. Right: virial masses of the MW analog and M31 (top and bottom, respectively).}
    \label{fig:M31}
\end{figure*}

Three trajectory variants taken from \cite{Arakelyan25} for three Local Group analogs from HESTIA numerical models (realizations 09\_18, 17\_11, 37\_17) are considered. HESTIA has 13 realizations differing by random parts in initial conditions used to fill data gaps. However, in some realizations, differences are only due to small-scale perturbations, and M31 and MW positions nearly coincide, so only three trajectory variants are presented in HESTIA data and considered here. Figure~\ref{fig:M31} shows M31 trajectories relative to the MW center. In all HESTIA realizations, the present-day radial and tangential velocities of M31 fall within observational constraints at the time of model creation (see \cite{Libeskind20}), yet the trajectories differ significantly.

It should be noted that the gravitational potential and M31 trajectory in our model are not self-consistent, i.e., they are not solutions to the two-body problem in the potentials of these two bodies. {This can be seen in Fig.~\ref{fig:M31} (top left), where the measured trajectories and Keplerian trajectories calculated from the initial points of the trajectories are shown.} We do not consider this a model deficiency since, in reality, the motion of the two galaxies should be influenced by the surrounding dark matter distribution, which is accounted for in the HESTIA numerical models but not in the two-body problem solution for MW and M31.

We use an orbit calculation method similar to that in \cite{2020MNRAS.499.4793S,2024ApJ...977...23A}. When modeling the potential with AGAMA, the motion of M31 and the non-inertial reference frame connected with the MW are taken into account. The MW position relative to the system's center of mass is computed, and in addition to the two-body potentials, an effective potential related to the MW acceleration is added. Due to inaccuracies in determining galaxy centers in simulations, their coordinates have small noise, which greatly increases when taking second derivatives to find accelerations. {Therefore, each of the three galaxy coordinate components and their masses over time were approximated by fourth-degree polynomials. Fig.~\ref{fig:M31} shows these approximated trajectories. Note that masses change complexly and sometimes decrease. This is related to defining virial mass as the total mass of particles within a certain radius. Some halo satellites can exit and then return inside this radius, causing oscillations in the mass defined this way \cite{Diemer21}.} Orbit integration of particles was performed using the \texttt{orbit} method from AGAMA.

Initial conditions were set as a random distribution on a sphere of radius $r_0$, with velocities chosen so that without M31, particles would move on randomly oriented circular orbits. For each particle, the orientation of its orbital plane at the initial time was recorded. As a measure of M31 perturbation influence on a particle, we consider the final distance $|Z|$ from the particle to its initial orbital plane. The larger this distance, the more unstable can be the plane in which the particle initially moved. This characteristic was chosen instead of angular momentum (used in Section 2) because $|Z|$ can be directly applied to analyzing the "thin plane" of satellites.

\begin{figure*}
    \centering
    \includegraphics[width=\linewidth]{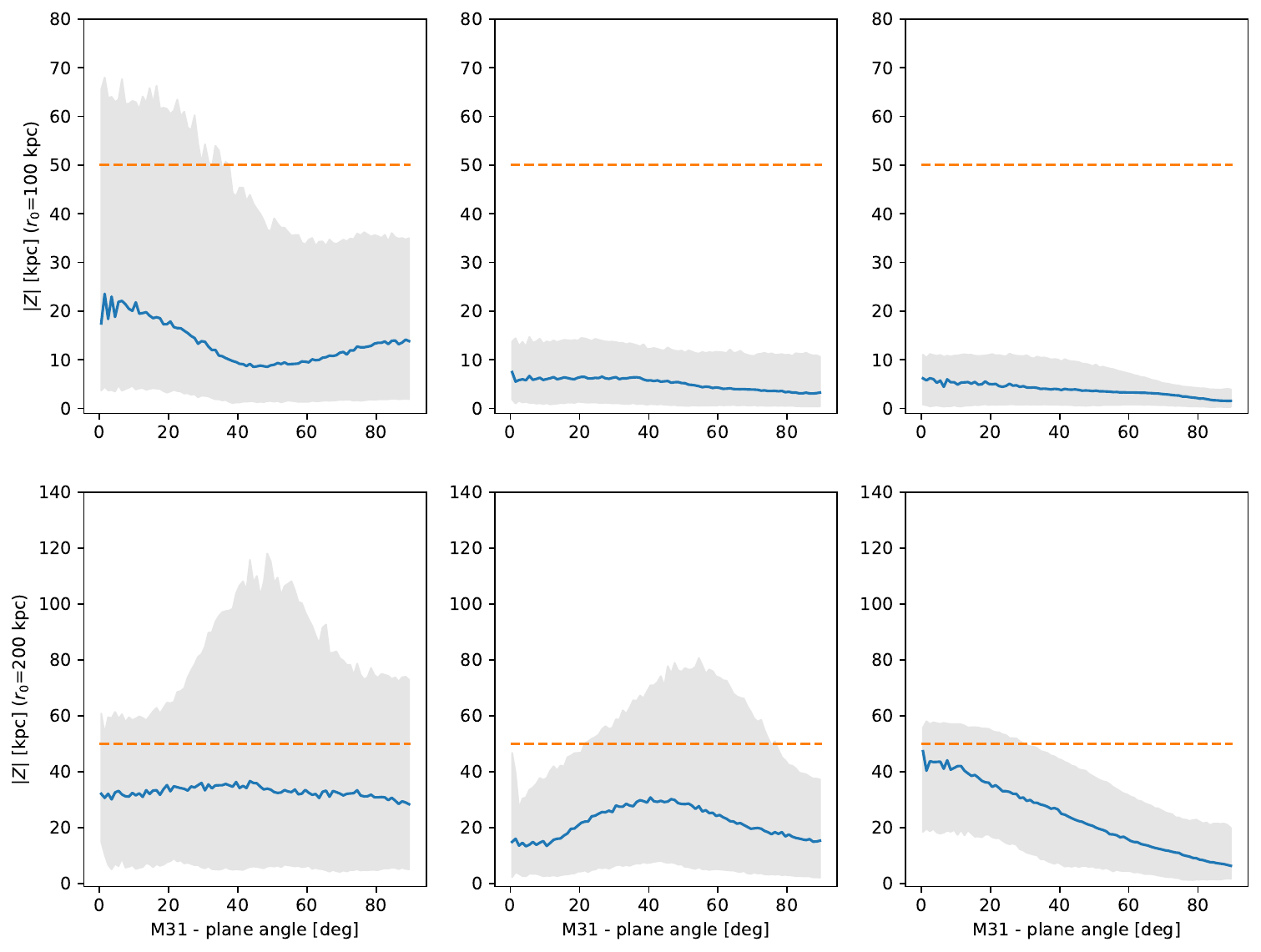}
    \caption{Distances of test particles from the initial plane after 6 billion years in three models with different M31 trajectories (left to right), depending on the angle between the plane's normal and the direction to M31. Solid lines show median distances, shading shows distances from 10\% to 90\% percentiles, dashed line shows 50 kpc, the thickness of the "thin plane." Top row: initial radius 100 kpc, bottom row: 200 kpc.}
    \label{fig:plane-dist}
\end{figure*}

Since the nonsphericity in our models is created solely by M31, stability dependence on the plane orientation relative to M31 direction can be expected. Figure~\ref{fig:plane-dist} shows final particle distances from the initial plane depending on the angle between the plane normal and the present-day direction to M31. Solid lines show medians, shading shows the range from the 10\% smallest to 10\% largest distances (10\% and 90\% percentiles).

The observed "thin plane" has a thickness of 50 kpc, which can be considered a threshold: if particles in the model have $|Z| > 50$ kpc, the satellite galaxy plane would not survive long. From Fig.~\ref{fig:plane-dist}, at initial radius 100 kpc, almost all particles deviate less than 50 kpc from the plane, i.e., the plane does not disperse. At 200 kpc, however, a significant fraction of particles are farther than 50 kpc from the initial plane.

\begin{figure*}
    \centering
    \includegraphics[width=\linewidth]{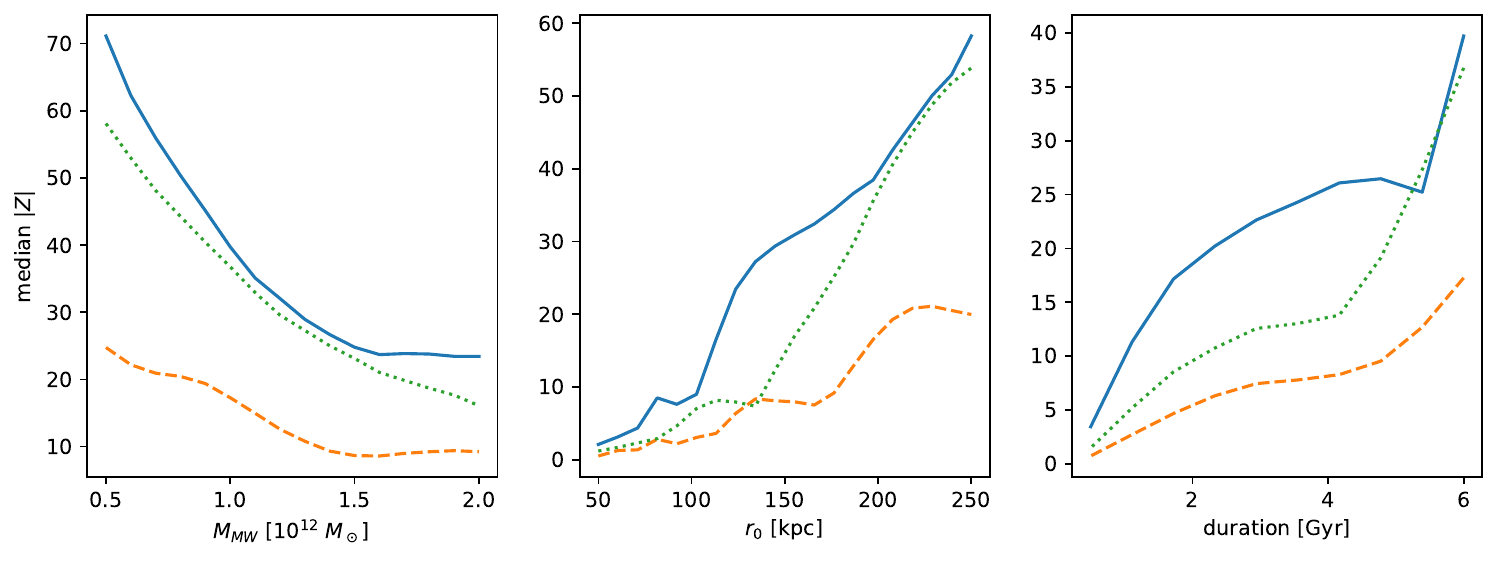}
    \caption{Median particle distance from the initial plane depending on the MW mass (left), initial orbit radius $r_0$ (center), and integration time (right). For left and right panels, $r_0=200$ kpc. Solid line — 09\_18, dashed — 17\_11, dotted — 37\_17.}
    \label{fig:analysis}
\end{figure*}

{There is a significant dependence on realization: the plane is most stable in realization 37\_17, where M31 mass is the smallest (see Fig.~\ref{fig:M31}).} For realizations 09\_18 and 37\_11, some dependence of plane stability on the angle between the plane normal and the current M31 direction is observed, but this dependence varies between realizations and distances. In our real Local Group, this angle is about 35$^\circ$ according to \cite{Arakelyan18}.

{The differences between the three realizations seen in Fig.~\ref{fig:plane-dist} may be caused by differences in MW and M31 masses and trajectories. To investigate this, additional calculations were performed with fixed galaxy masses: M31 mass, better known, was fixed at $1.6\times10^{12}$ M$_\odot$ \cite{Makarov25}, and MW mass varied from $0.5\times10^{12}$ to $2.0\times10^{12}$ M$_\odot$. We consider such artificial mass changes feasible since HESTIA trajectories differ significantly from Keplerian ones, indicating that the motion of MW and M31 is largely determined by the surrounding matter potential.}

The left panel of Fig.~\ref{fig:analysis} shows particle distances from the initial plane depending on MW mass for the three trajectory realizations. Both mass and trajectory strongly affect the result. Fig.~\ref{fig:analysis} also shows plane stability analysis results for fixed galaxy masses $M_{MW}=10^{12}$ M$_\odot$, $M_{M31}=1.6\times10^{12}$ M$_\odot$. The radius dependence shows good orbital plane stability at distances less than 75 kpc. In this case, the median distance from the plane is less than 10\% of the initial radius. This also gives an estimate of the error when reconstructing the orbit of some body (e.g., a GC) backward in time if M31 influence is neglected.

{In one realization, at distances greater than 125 kpc, the median distance from the plane reaches 50\% of the initial radius, and particles on circular orbits with radii greater than 150 kpc should leave a 100 kpc thick plane on average over 6 billion years.}

Most important, in our opinion, is the time dependence shown in the right panel of Fig.~\ref{fig:analysis}. At times less than 2 billion years, even for initial radius 200 kpc, the median distance is much less than 50 kpc, and almost all particles remain in the plane. At 6 billion years, about half of the particle trajectories exceed 50 kpc from the initial plane in realizations 09\_18 and 37\_17. This result is consistent with the idea that the observed satellites and GCs is a short-lived, temporary phenomenon.

\section{Discussion and Conclusions}
This work studies the motion of test bodies, initially located in one plane, in the gravitational potential of Galaxy analogs at significant distances from the center. The nonspherical and time-dependent potential was set based on results from the so-called "constrained" cosmological simulations HESTIA, which reproduce the large-scale structure of the observed Universe and contain Local Group analogs. In Section 2, the potential was modeled as an NFW profile perturbed by a quadrupole component with amplitude taken from simulations. In Section 3, the potential was modeled as a two-body system (MW and M31), with trajectories taken from numerical models. For simplicity, initial conditions for test particles were circular orbits in the unperturbed potential.

Since the real masses and potentials in the vicinity of the Galaxy are unknown, as is the M31 trajectory relative to MW {(see, e.g., \cite{Sawala25})}, we studied how orbit perturbations depend on model parameters. In Section 2, it is shown that exceeding a certain quadrupole amplitude (depending on orbit radius) leads to angular momentum changes over 5 billion years exceeding the initial angular momentum.

Section 3 examines the influence of the perturbation created by M31 on the orbits of the test bodies. This influence is measured by the distance $|Z|$ of a particle from its initial orbital plane at the end of the calculation. Main results are shown in Figs.~\ref{fig:plane-dist} and \ref{fig:analysis}. At distances greater than 150 kpc, M31 influence can cause particle trajectories to deviate by $|Z|>50$ kpc over 4-6 billion years, which would destroy the "thin plane" of satellite galaxies. However, this conclusion depends on the still unknown M31 trajectory and galaxy masses. At times less than 3 billion years, for all considered trajectories, most particles deviate much less than 50 kpc from the initial plane.

At distances where GCs exist, i.e., less than 100 kpc, deviations $|Z|$ are less than 10 kpc (and less than 10\% of the initial orbit radius) over 6 billion years. The time dependence of deviations averaged over many orbits is approximately linear. This gives an estimate of the error in reconstructing GC trajectories backward in time when neglecting M31 influence.

The authors thank the HESTIA project team: N. Libeskind, S. Gottl\"{o}ber, G. Yepes, Y. Hoffman, A. Knebe, and others for providing data.

\bibliographystyle{ieeetr}
\bibliography{wave.bib}

@ARTICLE{Makarov25,
       author = {{Makarov}, Danila and {Makarov}, Dmitry and {Kozyrev}, Kirill and {Libeskind}, Noam},
        title = "{Line-of-Sight Mass Estimator and the Masses of the Milky Way and Andromeda Galaxy}",
      journal = {Universe},
         year = 2025,
        month = apr,
       volume = {11},
       number = {5},
          eid = {144},
        pages = {144},
          doi = {10.3390/universe11050144},
archivePrefix = {arXiv},
       eprint = {2503.12612},
 primaryClass = {astro-ph.GA},
       adsurl = {https://ui.adsabs.harvard.edu/abs/2025Univ...11..144M}
}

@ARTICLE{Diemer21,
       author = {{Diemer}, Benedikt},
        title = "{Flybys, Orbits, Splashback: Subhalos and the Importance of the Halo Boundary}",
      journal = {\apj},
         year = 2021,
        month = mar,
       volume = {909},
       number = {2},
          eid = {112},
        pages = {112},
          doi = {10.3847/1538-4357/abd947},
archivePrefix = {arXiv},
       eprint = {2007.10992},
 primaryClass = {astro-ph.CO},
       adsurl = {https://ui.adsabs.harvard.edu/abs/2021ApJ...909..112D}
}

@ARTICLE{Sawala25,
       author = {{Sawala}, Till and {Delhomelle}, Jehanne and {Deason}, Alis J. and {Frenk}, Carlos S. and {H{\"a}kkinen}, Jenni and {Johansson}, Peter H. and {Keitaanranta}, Atte and {Rawlings}, Alexander and {Wright}, Ruby},
        title = "{No certainty of a Milky Way-Andromeda collision}",
      journal = {Nature Astronomy},
         year = 2025,
        month = aug,
       volume = {9},
        pages = {1206-1217},
          doi = {10.1038/s41550-025-02563-1},
       adsurl = {https://ui.adsabs.harvard.edu/abs/2025NatAs...9.1206S}
}

@ARTICLE{Samuel21,
       author = {{Samuel}, Jenna and {Wetzel}, Andrew and {Chapman}, Sierra and {Tollerud}, Erik and {Hopkins}, Philip F. and {Boylan-Kolchin}, Michael and {Bailin}, Jeremy and {Faucher-Gigu{\`e}re}, Claude-Andr{\'e}},
        title = "{Planes of satellites around Milky Way/M31-mass galaxies in the FIRE simulations and comparisons with the Local Group}",
      journal = {\mnras},
         year = 2021,
        month = jun,
       volume = {504},
       number = {1},
        pages = {1379-1397},
          doi = {10.1093/mnras/stab955},
archivePrefix = {arXiv},
       eprint = {2010.08571},
 primaryClass = {astro-ph.GA},
       adsurl = {https://ui.adsabs.harvard.edu/abs/2021MNRAS.504.1379S}
}

@ARTICLE{Ivanov25,
       author = {{Ivanov}, Pavel B. and {Polnarev}, Alexander G.},
        title = "{The evolution of a supermassive binary black hole in an non-spherical nuclear star cluster}",
      journal = {arXiv e-prints},
         year = 2025,
        month = jul,
          eid = {arXiv:2507.11684},
        pages = {arXiv:2507.11684},
          doi = {10.48550/arXiv.2507.11684},
archivePrefix = {arXiv},
       eprint = {2507.11684},
 primaryClass = {astro-ph.GA},
       adsurl = {https://ui.adsabs.harvard.edu/abs/2025arXiv250711684I}
}

@ARTICLE{SLOW,
       author = {{Dolag}, Klaus and {Sorce}, Jenny G. and {Pilipenko}, Sergey and {Hern{\'a}ndez-Mart{\'\i}nez}, Elena and {Valentini}, Milena and {Gottl{\"o}ber}, Stefan and {Aghanim}, Nabila and {Khabibullin}, Ildar},
        title = "{Simulating the LOcal Web (SLOW). I. Anomalies in the local density field}",
      journal = {\aap},
         year = 2023,
        month = sep,
       volume = {677},
          eid = {A169},
        pages = {A169},
          doi = {10.1051/0004-6361/202346213},
archivePrefix = {arXiv},
       eprint = {2302.10960},
 primaryClass = {astro-ph.CO},
       adsurl = {https://ui.adsabs.harvard.edu/abs/2023A&A...677A.169D}
}

@ARTICLE{2015MNRAS.452.1052L,
       author = {{Libeskind}, Noam I. and {Hoffman}, Yehuda and {Tully}, R. Brent and {Courtois}, Helene M. and {Pomar{\`e}de}, Daniel and {Gottl{\"o}ber}, Stefan and {Steinmetz}, Matthias},
        title = "{Planes of satellite galaxies and the cosmic web}",
      journal = {\mnras},
         year = 2015,
        month = sep,
       volume = {452},
       number = {1},
        pages = {1052-1059},
          doi = {10.1093/mnras/stv1302},
archivePrefix = {arXiv},
       eprint = {1503.05915},
 primaryClass = {astro-ph.GA},
       adsurl = {https://ui.adsabs.harvard.edu/abs/2015MNRAS.452.1052L}
}

@ARTICLE{CLUES,
       author = {{Carlesi}, Edoardo and {Sorce}, Jenny G. and {Hoffman}, Yehuda and {Gottl{\"o}ber}, Stefan and {Yepes}, Gustavo and {Libeskind}, Noam I. and {Pilipenko}, Sergey V. and {Knebe}, Alexander and {Courtois}, H{\'e}l{\`e}ne and {Tully}, R. Brent and {Steinmetz}, Matthias},
        title = "{Constrained Local UniversE Simulations: a Local Group factory}",
      journal = {\mnras},
         year = 2016,
        month = may,
       volume = {458},
       number = {1},
        pages = {900-911},
          doi = {10.1093/mnras/stw357},
archivePrefix = {arXiv},
       eprint = {1602.03919},
 primaryClass = {astro-ph.CO},
       adsurl = {https://ui.adsabs.harvard.edu/abs/2016MNRAS.458..900C}
}

@ARTICLE{NFW,
       author = {{Navarro}, Julio F. and {Frenk}, Carlos S. and {White}, Simon D.~M.},
        title = "{The Structure of Cold Dark Matter Halos}",
      journal = {\apj},
         year = 1996,
        month = may,
       volume = {462},
        pages = {563},
          doi = {10.1086/177173},
archivePrefix = {arXiv},
       eprint = {astro-ph/9508025},
 primaryClass = {astro-ph},
       adsurl = {https://ui.adsabs.harvard.edu/abs/1996ApJ...462..563N}
}

@ARTICLE{2015MNRAS.453.1047P,
       author = {{Pawlowski}, Marcel S. and {McGaugh}, Stacy S. and {Jerjen}, Helmut},
        title = "{The new Milky Way satellites: alignment with the VPOS and predictions for proper motions and velocity dispersions}",
      journal = {\mnras},
         year = 2015,
        month = oct,
       volume = {453},
       number = {1},
        pages = {1047-1061},
          doi = {10.1093/mnras/stv1588},
archivePrefix = {arXiv},
       eprint = {1505.07465},
 primaryClass = {astro-ph.GA},
       adsurl = {https://ui.adsabs.harvard.edu/abs/2015MNRAS.453.1047P}
}

@ARTICLE{2009MNRAS.394.2223M,
       author = {{Metz}, Manuel and {Kroupa}, Pavel and {Jerjen}, Helmut},
        title = "{Discs of satellites: the new dwarf spheroidals}",
      journal = {\mnras},
         year = 2009,
        month = apr,
       volume = {394},
       number = {4},
        pages = {2223-2228},
          doi = {10.1111/j.1365-2966.2009.14489.x},
archivePrefix = {arXiv},
       eprint = {0901.1658},
 primaryClass = {astro-ph.GA},
       adsurl = {https://ui.adsabs.harvard.edu/abs/2009MNRAS.394.2223M}
}

@ARTICLE{2007MNRAS.374.1125M,
       author = {{Metz}, Manuel and {Kroupa}, Pavel and {Jerjen}, Helmut},
        title = "{The spatial distribution of the Milky Way and Andromeda satellite galaxies}",
      journal = {\mnras},
         year = 2007,
        month = jan,
       volume = {374},
       number = {3},
        pages = {1125-1145},
          doi = {10.1111/j.1365-2966.2006.11228.x},
archivePrefix = {arXiv},
       eprint = {astro-ph/0610933},
 primaryClass = {astro-ph},
       adsurl = {https://ui.adsabs.harvard.edu/abs/2007MNRAS.374.1125M}
}

@ARTICLE{2008ApJ...680..287M,
       author = {{Metz}, Manuel and {Kroupa}, Pavel and {Libeskind}, Noam I.},
        title = "{The Orbital Poles of Milky Way Satellite Galaxies: A Rotationally Supported Disk of Satellites}",
      journal = {\apj},
         year = 2008,
        month = jun,
       volume = {680},
       number = {1},
        pages = {287-294},
          doi = {10.1086/587833},
archivePrefix = {arXiv},
       eprint = {0802.3899},
 primaryClass = {astro-ph},
       adsurl = {https://ui.adsabs.harvard.edu/abs/2008ApJ...680..287M}
}

@ARTICLE{2005A&A...431..517K,
       author = {{Kroupa}, P. and {Theis}, C. and {Boily}, C.~M.},
        title = "{The great disk of Milky-Way satellites and cosmological sub-structures}",
      journal = {\aap},
         year = 2005,
        month = feb,
       volume = {431},
        pages = {517-521},
          doi = {10.1051/0004-6361:20041122},
archivePrefix = {arXiv},
       eprint = {astro-ph/0410421},
 primaryClass = {astro-ph},
       adsurl = {https://ui.adsabs.harvard.edu/abs/2005A&A...431..517K}
}

@ARTICLE{1995MNRAS.275..429L,
       author = {{Lynden-Bell}, D. and {Lynden-Bell}, R.~M.},
        title = "{Ghostly streams from the formation of the Galaxy's halo}",
      journal = {\mnras},
         year = 1995,
        month = jul,
       volume = {275},
       number = {2},
        pages = {429-442},
          doi = {10.1093/mnras/275.2.429},
       adsurl = {https://ui.adsabs.harvard.edu/abs/1995MNRAS.275..429L}
}

@ARTICLE{1976MNRAS.174..695L,
       author = {{Lynden-Bell}, D.},
        title = "{Dwarf galaxies and globular clusters in high velocity hydrogen streams.}",
      journal = {\mnras},
         year = 1976,
        month = mar,
       volume = {174},
        pages = {695-710},
          doi = {10.1093/mnras/174.3.695},
       adsurl = {https://ui.adsabs.harvard.edu/abs/1976MNRAS.174..695L}
}

@ARTICLE{2025Univ...11..191M,
       author = {{Machado}, Rubens E.~G. and {Tauil}, Giovanni C. and {Schweder-Souza}, Nicholas},
        title = "{Accuracy of Analytic Potentials for Orbits of Satellites Around a Milky Way-like Galaxy: Comparison with N-Body Simulations}",
      journal = {Universe},
         year = 2025,
        month = jun,
       volume = {11},
       number = {6},
          eid = {191},
        pages = {191},
          doi = {10.3390/universe11060191},
archivePrefix = {arXiv},
       eprint = {2506.13813},
 primaryClass = {astro-ph.GA},
       adsurl = {https://ui.adsabs.harvard.edu/abs/2025Univ...11..191M}
}

@ARTICLE{2025arXiv250601459K,
       author = {{Kumar}, Prem and {Pawlowski}, Marcel S. and {Kanehisa}, Kosuke Jamie and {Li}, Pengfei and {J{\'u}lio}, Mariana P. and {Taibi}, Salvatore},
        title = "{The effect of measurement uncertainties on the inferred stability of planes of satellite galaxies}",
      journal = {arXiv e-prints},
         year = 2025,
        month = jun,
          eid = {arXiv:2506.01459},
        pages = {arXiv:2506.01459},
          doi = {10.48550/arXiv.2506.01459},
archivePrefix = {arXiv},
       eprint = {2506.01459},
 primaryClass = {astro-ph.GA},
       adsurl = {https://ui.adsabs.harvard.edu/abs/2025arXiv250601459K}
}

@ARTICLE{2021ApJ...923..140G,
       author = {{Garavito-Camargo}, Nicol{\'a}s and {Patel}, Ekta and {Besla}, Gurtina and {Price-Whelan}, Adrian M. and {G{\'o}mez}, Facundo A. and {Laporte}, Chervin F.~P. and {Johnston}, Kathryn V.},
        title = "{The Clustering of Orbital Poles Induced by the LMC: Hints for the Origin of Planes of Satellites}",
      journal = {\apj},
         year = 2021,
        month = dec,
       volume = {923},
       number = {2},
          eid = {140},
        pages = {140},
          doi = {10.3847/1538-4357/ac2c05},
archivePrefix = {arXiv},
       eprint = {2108.07321},
 primaryClass = {astro-ph.GA},
       adsurl = {https://ui.adsabs.harvard.edu/abs/2021ApJ...923..140G}
}

@ARTICLE{2022A&A...657A..54B,
       author = {{Battaglia}, G. and {Taibi}, S. and {Thomas}, G.~F. and {Fritz}, T.~K.},
        title = "{Gaia early DR3 systemic motions of Local Group dwarf galaxies and orbital properties with a massive Large Magellanic Cloud}",
      journal = {\aap},
         year = 2022,
        month = jan,
       volume = {657},
          eid = {A54},
        pages = {A54},
          doi = {10.1051/0004-6361/202141528},
archivePrefix = {arXiv},
       eprint = {2106.08819},
 primaryClass = {astro-ph.GA},
       adsurl = {https://ui.adsabs.harvard.edu/abs/2022A&A...657A..54B}
}

@ARTICLE{2022ApJ...940..136P,
       author = {{Pace}, Andrew B. and {Erkal}, Denis and {Li}, Ting S.},
        title = "{Proper Motions, Orbits, and Tidal Influences of Milky Way Dwarf Spheroidal Galaxies}",
      journal = {\apj},
         year = 2022,
        month = dec,
       volume = {940},
       number = {2},
          eid = {136},
        pages = {136},
          doi = {10.3847/1538-4357/ac997b},
archivePrefix = {arXiv},
       eprint = {2205.05699},
 primaryClass = {astro-ph.GA},
       adsurl = {https://ui.adsabs.harvard.edu/abs/2022ApJ...940..136P}
}

@ARTICLE{2022ARep...66..359A,
       author = {{Arakelyan}, N.~R. and {Pilipenko}, S.~V.},
        title = "{Erratum to: Globular Cluster as Indicators of Galactic Evolution}",
      journal = {Astronomy Reports},
         year = 2022,
        month = apr,
       volume = {66},
       number = {4},
        pages = {359-359},
          doi = {10.1134/S1063772922330010},
archivePrefix = {arXiv},
       eprint = {2301.08535},
 primaryClass = {astro-ph.GA},
       adsurl = {https://ui.adsabs.harvard.edu/abs/2022ARep...66..359A}
}

@ARTICLE{2025A&A...694A..66A,
       author = {{Akib}, Istiak and {Hammer}, Fran{\c{c}}ois and {Yang}, Yanbin and {Pawlowski}, Marcel S. and {Wang}, Jianling},
        title = "{An intriguing coincidence between the majority of vast polar structure dwarfs and a recent major merger at the M31 position}",
      journal = {\aap},
         year = 2025,
        month = feb,
       volume = {694},
          eid = {A66},
        pages = {A66},
          doi = {10.1051/0004-6361/202452131},
archivePrefix = {arXiv},
       eprint = {2501.00870},
 primaryClass = {astro-ph.GA},
       adsurl = {https://ui.adsabs.harvard.edu/abs/2025A&A...694A..66A}
}

@ARTICLE{2021ApJ...916....8L,
       author = {{Li}, Hefan and {Hammer}, Francois and {Babusiaux}, Carine and {Pawlowski}, Marcel S. and {Yang}, Yanbin and {Arenou}, Frederic and {Du}, Cuihua and {Wang}, Jianling},
        title = "{Gaia EDR3 Proper Motions of Milky Way Dwarfs. I. 3D Motions and Orbits}",
      journal = {\apj},
         year = 2021,
        month = jul,
       volume = {916},
       number = {1},
          eid = {8},
        pages = {8},
          doi = {10.3847/1538-4357/ac0436},
archivePrefix = {arXiv},
       eprint = {2104.03974},
 primaryClass = {astro-ph.GA},
       adsurl = {https://ui.adsabs.harvard.edu/abs/2021ApJ...916....8L}
}

@ARTICLE{2024A&A...681A..73T,
       author = {{Taibi}, S. and {Pawlowski}, M.~S. and {Khoperskov}, S. and {Steinmetz}, M. and {Libeskind}, N.~I.},
        title = "{A portrait of the vast polar structure as a young phenomenon: Hints from its member satellites}",
      journal = {\aap},
         year = 2024,
        month = jan,
       volume = {681},
          eid = {A73},
        pages = {A73},
          doi = {10.1051/0004-6361/202347473},
archivePrefix = {arXiv},
       eprint = {2310.13521},
 primaryClass = {astro-ph.GA},
       adsurl = {https://ui.adsabs.harvard.edu/abs/2024A&A...681A..73T}
}

@ARTICLE{Arakelyan25,
       author = {{Arakelyan}, Naira R. and {Pilipenko}, Sergey V. and {Gottl{\"o}ber}, Stefan and {Libeskind}, Noam I. and {Yepes}, Gustavo and {Hoffman}, Yehuda},
        title = "{Variable gravitational potential of Milky Way analogues in HESTIA suite}",
      journal = {arXiv e-prints},
         year = 2024,
        month = dec,
          eid = {arXiv:2412.18880},
        pages = {arXiv:2412.18880},
          doi = {10.48550/arXiv.2412.18880},
archivePrefix = {arXiv},
       eprint = {2412.18880},
 primaryClass = {astro-ph.GA},
       adsurl = {https://ui.adsabs.harvard.edu/abs/2024arXiv241218880A}
}

@ARTICLE{2022ApJ...933..113R,
       author = {{Rozier}, Simon and {Famaey}, Benoit and {Siebert}, Arnaud and {Monari}, Giacomo and {Pichon}, Christophe and {Ibata}, Rodrigo},
        title = "{Constraining the Milky Way Halo Kinematics via Its Linear Response to the Large Magellanic Cloud}",
      journal = {\apj},
         year = 2022,
        month = jul,
       volume = {933},
       number = {1},
          eid = {113},
        pages = {113},
          doi = {10.3847/1538-4357/ac7139},
archivePrefix = {arXiv},
       eprint = {2201.05589},
 primaryClass = {astro-ph.GA},
       adsurl = {https://ui.adsabs.harvard.edu/abs/2022ApJ...933..113R}
}

@ARTICLE{2025ApJ...978...79B,
       author = {{Brooks}, Richard A.~N. and {Garavito-Camargo}, Nicol{\'a}s and {Johnston}, Kathryn V. and {Price-Whelan}, Adrian M. and {Sanders}, Jason L. and {Lilleengen}, Sophia},
        title = "{LMC Calls, Milky Way Halo Answers: Disentangling the Effects of the MW{\textendash}LMC Interaction on Stellar Stream Populations}",
      journal = {\apj},
         year = 2025,
        month = jan,
       volume = {978},
       number = {1},
          eid = {79},
        pages = {79},
          doi = {10.3847/1538-4357/ad93a7},
archivePrefix = {arXiv},
       eprint = {2410.02574},
 primaryClass = {astro-ph.GA},
       adsurl = {https://ui.adsabs.harvard.edu/abs/2025ApJ...978...79B}
}

@ARTICLE{CorreaMagnus22,
       author = {{Correa Magnus}, Lilia and {Vasiliev}, Eugene},
        title = "{Measuring the Milky Way mass distribution in the presence of the LMC}",
      journal = {\mnras},
         year = 2022,
        month = apr,
       volume = {511},
       number = {2},
        pages = {2610-2630},
          doi = {10.1093/mnras/stab3726},
archivePrefix = {arXiv},
       eprint = {2110.00018},
 primaryClass = {astro-ph.GA},
       adsurl = {https://ui.adsabs.harvard.edu/abs/2022MNRAS.511.2610C}
}

@ARTICLE{Libeskind20,
       author = {{Libeskind}, Noam I. and {Carlesi}, Edoardo and {Grand}, Robert J.~J. and {Khalatyan}, Arman and {Knebe}, Alexander and {Pakmor}, Ruediger and {Pilipenko}, Sergey and {Pawlowski}, Marcel S. and {Sparre}, Martin and {Tempel}, Elmo and {Wang}, Peng and {Courtois}, H{\'e}l{\`e}ne M. and {Gottl{\"o}ber}, Stefan and {Hoffman}, Yehuda and {Minchev}, Ivan and {Pfrommer}, Christoph and {Sorce}, Jenny G. and {Springel}, Volker and {Steinmetz}, Matthias and {Tully}, R. Brent and {Vogelsberger}, Mark and {Yepes}, Gustavo},
        title = "{The HESTIA project: simulations of the Local Group}",
      journal = {\mnras},
         year = 2020,
        month = oct,
       volume = {498},
       number = {2},
        pages = {2968-2983},
          doi = {10.1093/mnras/staa2541},
archivePrefix = {arXiv},
       eprint = {2008.04926},
 primaryClass = {astro-ph.GA},
       adsurl = {https://ui.adsabs.harvard.edu/abs/2020MNRAS.498.2968L}
}

@ARTICLE{Makarov23,
       author = {{Makarov}, Dmitry and {Khoperskov}, Sergey and {Makarov}, Danila and {Makarova}, Lidia and {Libeskind}, Noam and {Salomon}, Jean-Baptiste},
        title = "{The LMC impact on the kinematics of the Milky Way satellites: clues from the running solar apex}",
      journal = {\mnras},
         year = 2023,
        month = may,
       volume = {521},
       number = {3},
        pages = {3540-3552},
          doi = {10.1093/mnras/stad757},
archivePrefix = {arXiv},
       eprint = {2303.06175},
 primaryClass = {astro-ph.GA},
       adsurl = {https://ui.adsabs.harvard.edu/abs/2023MNRAS.521.3540M}
}

@ARTICLE{Arakelyan22,
       author = {{Arakelyan}, N.~R. and {Pilipenko}, S.~V.},
        title = "{Erratum to: Globular Cluster as Indicators of Galactic Evolution}",
      journal = {Astronomy Reports},
         year = 2022,
        month = apr,
       volume = {66},
       number = {4},
        pages = {359-359},
          doi = {10.1134/S1063772922330010},
archivePrefix = {arXiv},
       eprint = {2301.08535},
 primaryClass = {astro-ph.GA},
       adsurl = {https://ui.adsabs.harvard.edu/abs/2022ARep...66..359A}
}

@ARTICLE{Arakelyan18,
       author = {{Arakelyan}, N.~R. and {Pilipenko}, S.~V. and {Libeskind}, N.~I.},
        title = "{Spatial distribution of globular clusters in the Galaxy}",
      journal = {\mnras},
         year = 2018,
        month = nov,
       volume = {481},
       number = {1},
        pages = {918-929},
          doi = {10.1093/mnras/sty2320},
archivePrefix = {arXiv},
       eprint = {1803.04770},
 primaryClass = {astro-ph.GA},
       adsurl = {https://ui.adsabs.harvard.edu/abs/2018MNRAS.481..918A}
}

@ARTICLE{2018MNRAS.473.2212F,
       author = {{Fernando}, Nuwanthika and {Arias}, Veronica and {Lewis}, Geraint F. and {Ibata}, Rodrigo A. and {Power}, Chris},
        title = "{Stability of satellite planes in M31 II: effects of the dark subhalo population}",
      journal = {\mnras},
         year = 2018,
        month = jan,
       volume = {473},
       number = {2},
        pages = {2212-2221},
          doi = {10.1093/mnras/stx2483},
archivePrefix = {arXiv},
       eprint = {1709.08010},
 primaryClass = {astro-ph.GA},
       adsurl = {https://ui.adsabs.harvard.edu/abs/2018MNRAS.473.2212F}
}

@ARTICLE{2018arXiv180208255V,
       author = {{Vasiliev}, Eugene},
        title = "{Agama reference documentation}",
      journal = {arXiv e-prints},
         year = 2018,
        month = feb,
          eid = {arXiv:1802.08255},
        pages = {arXiv:1802.08255},
          doi = {10.48550/arXiv.1802.08255},
archivePrefix = {arXiv},
       eprint = {1802.08255},
 primaryClass = {astro-ph.IM},
       adsurl = {https://ui.adsabs.harvard.edu/abs/2018arXiv180208255V}
}

@ARTICLE{2020MNRAS.499.4793S,
       author = {{Sanders}, Jason L. and {Lilley}, Edward J. and {Vasiliev}, Eugene and {Evans}, N. Wyn and {Erkal}, Denis},
        title = "{Models of distorted and evolving dark matter haloes}",
      journal = {\mnras},
         year = 2020,
        month = dec,
       volume = {499},
       number = {4},
        pages = {4793-4813},
          doi = {10.1093/mnras/staa3079},
archivePrefix = {arXiv},
       eprint = {2009.00645},
 primaryClass = {astro-ph.GA},
       adsurl = {https://ui.adsabs.harvard.edu/abs/2020MNRAS.499.4793S}
}

@ARTICLE{2023Galax..11...59V,
       author = {{Vasiliev}, Eugene},
        title = "{The Effect of the LMC on the Milky Way System}",
      journal = {Galaxies},
         year = 2023,
        month = apr,
       volume = {11},
       number = {2},
          eid = {59},
        pages = {59},
          doi = {10.3390/galaxies11020059},
archivePrefix = {arXiv},
       eprint = {2304.09136},
 primaryClass = {astro-ph.GA},
       adsurl = {https://ui.adsabs.harvard.edu/abs/2023Galax..11...59V}
}

@ARTICLE{2024ApJ...977...23A,
       author = {{Arora}, Arpit and {Sanderson}, Robyn and {Regan}, Christopher and {Garavito-Camargo}, Nicol{\'a}s and {Bregou}, Emily and {Panithanpaisal}, Nondh and {Wetzel}, Andrew and {Cunningham}, Emily C. and {Loebman}, Sarah R. and {Dropulic}, Adriana and {Shipp}, Nora},
        title = "{Efficient and Accurate Force Replay in Cosmological-baryonic Simulations}",
      journal = {\apj},
         year = 2024,
        month = dec,
       volume = {977},
       number = {1},
          eid = {23},
        pages = {23},
          doi = {10.3847/1538-4357/ad88f0},
archivePrefix = {arXiv},
       eprint = {2407.12932},
 primaryClass = {astro-ph.GA},
       adsurl = {https://ui.adsabs.harvard.edu/abs/2024ApJ...977...23A}
}

\end{document}